\documentstyle[psfig]{l-aa}

\def\vvmm{$\langle V / V_{max}  \rangle$}

\begin{document}

   \thesaurus{13         
              (13.07.1)} 
   \title{Gamma-ray bursts : towards a standard candle luminosity}

   \subtitle{}

   \author{J-L. Atteia
          }

   \offprints{J-L. Atteia}

   \institute{Centre d'Etude Spatiale des Rayonnements (CNRS/UPS), 
              BP 4346, 31028 Toulouse Cedex 4, France\\ 
              email: atteia@cesr.cnes.fr
             }

   \date{Received ?? ; accepted ??}

   \maketitle

   \begin{abstract}
   It is usual, in gamma-ray burst (GRB) studies, to compare the average
properties of bright and faint GRBs, with the assumption that 
brightness classes reflect distance classes.
When brightness is intented to reflect the distance to the sources,
it is nevertheless important to use a quantity with a small intrinsic
dispersion.
We propose here a method to compare the intrinsic dispersion of various
measures of GRB brightness. 
This method assumes that nearby bursters are homogeneously distributed
in an Euclidean space with no density or luminosity evolution.
We then use it to compare 5 measures of GRB brightness 
in the BATSE Catalog.
Our analysis reveals that better (i.e. less dispersed) measures
of brightness are obtained at low energy and that GRBs 
are much closer to standard candles below 100 keV than above.
We suggest that a beaming of the emission above 100~keV could explain 
this behaviour. 

\keywords{Gamma-rays : bursts}
   \end{abstract}


\section{Introduction}

The GRB intensity distribution has been extensively used 
to improve our understanding of these sources.
The \vvmm\ test for instance has been unvaluable to demonstrate
the burster spatial inhomogeneity (Meegan et al. 1992).
The LogN-LogP distribution has been shown to be in good agreement
with the intensity distribution expected for cosmological sources 
(e.g. Piran 1992, Wickramasinghe et al. 1993, Fenimore et al. 1993).   
Finally, it is common to define brightness classes for the purpose
of searching cosmological signatures in gamma-ray bursts.
In all these studies, the burst brightness appears as a key parameter.

While it is widely acknowledged that C$_{max}$/C$_{min}$ is the
measure of brightness which is appropriate to check the spatial 
homogeneity of gamma-ray bursters (Schmidt et al. 1988), 
there is no such agreement on the parameter which should be used 
for studies where brightness is taken as a distance indicator.
These studies require a quantity with small, or no, intrinsic dispersion.
We propose below a new way of comparing the 
intrinsic dispersions of different measures of brightness.
When several definitions of brightness are available 
(e.g. peak flux, fluence...), this method answers the question 
of which one of these quantities is closer to a standard candle.
Section 2 describes the method.
In section 3, we apply it to a sample of 1471 GRBs detected by BATSE.
The implications for gamma-ray bursts are discussed in section 4.

\section{The method}


We propose a method based on the observation that the sources
detected by an instrument have different spatial distributions
depending on whether they are standard candles or not.
Standard candles are fully sampled out to a distance $D_0$ and not seen
beyond this distance (assuming we have a detector with a sharp intensity 
threshold). 
Sources with a broad luminosity function, on the other hand, 
are fully sampled out to a distance $D_1$
(the distance at which the intensity of intrinsically faint sources 
reaches the threshold of the instrument), and only partially
detected between $D_1$ and $D_2$ (the distance at which the intensity 
of intrinsically bright sources reaches the sensitivity threshold 
of the instrument).

Let us consider now the case of a population 
which is homogeneous for  $D < D_h$, 
and whose spatial density decreases for $D > D_h$. 
For standard candles, the deviation from homogeneity in the 
size-frequency distribution cannot be seen until {\it all the sources 
in the homogeneous region have been detected}.
For sources with a broad luminosity distribution, the deviation
from homogeneity is in principle visible when the instrument
begins to detect sources in the non-homogeneous region.
If the luminosity function is broad enough, this happens while most 
of the sources in the homogeneous region are still undetected. 
In this second situation, the number of sources seen 
in the homogeneous part of the observed size-frequency relation 
can be significantly smaller than for standard candles.

Figure~1 illustrates this behaviour for sources having 
a power law luminosity distribution 
($n(L) \propto L^{-\alpha}$ between $L_1$ and $L_2$).
This figure displays $f_{3/2}$ as a function of the ratio 
$L_2 / L_1$, where $f_{3/2}$ is
the ratio of the number of sources {\it seen} in the homogeneous 
part of the size-frequency curve normalized to the same number
for standard candles.
As expected, $f_{3/2}$ continuously
decreases when the luminosity dispersion increases.

\subsection{Using the GRB size-frequency distribution}

As explained above, better measures of brightness are indicated
by a larger number of GRBs in the homogeneous part of their
size-frequency curve. 
This is the criterion we propose to use to compare the "quality" 
of different measures of GRB brightness.

Practically, we measure the point at which the observed size-frequency
curve deviates from the extrapolation of the bright end 
of the distribution.
In order to decrease the sensitivity of our test to the behaviour
of the curve below the break (which depends on the source luminosity
function and on their radial density profile), we have chosen 
to count the number of GRBs, $N_b$, when the observed curve is only 20\%
below the extrapolation of the "homogeneous" part of the curve
(i.e. log(N$_{\rm extrapolated}$) - log(N$_{\rm observed}$) = 0.1).
The confidence region on  $N_b$ is computed with 
the bootstrap method.
We construct several simulated samples from the original sample,
and we compute a new value of $N_b$ for each of them.
The confidence region is defined as the interval which contains
a given percentage of the simulated $N_b$.
The error bars on $N_b$ are given below at the 90\% confidence level.

As we do not know $D_h$, nor the number of bursters in the
homogeneous region, this method can only tell us that one brightness
indicator is better than another, but not whether it is a true
standard candle.

   \begin{figure}
    \psfig{file=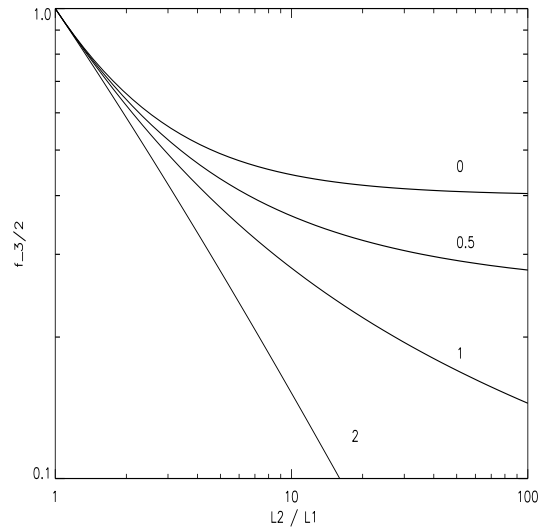,height=7.5cm,width=7.5cm}
     \caption{Influence of an intrinsic luminosity distribution on the number
of GRBs seen in the homogeneous part of the size-frequency curve.
We assume that the luminosity function has the form 
$n(L) \propto L^{- \alpha}$ between $L_1$ and $L_2$.
The figure displays $f_{3/2}$ as a function of $L_2 / L_1$, where
$f_{3/2}$ is the ratio of the number of GRBs seen in the homogeneous
part of the size-frequency curve to the number expected for 
standard candles. Note that $f_{3/2}$ decreases when $L_2 / L_1$ 
increases. The curves are for $\alpha = 0, 0.5, 1, 2$ (top to bottom).}
         \label{rapn}
    \end{figure}

\subsection{Limits of the method}

Our analysis relies on the following assumptions :

1) The slope -3/2 observed at the bright end of the size-frequency
distribution is actually due to the spatial homogeneity of nearby 
bursters.

2) BATSE is sensitive enough to detect all the bursts emitted inside 
the homogeneous region (we show below that this second assumption is
fully justified).

Under these assumptions, the results presented below are valid
for both cosmological and galactic models.
It should be noted however that this analysis can only be applied 
to non-evolving sources in an Euclidean space.
If gamma-ray bursters appear to be at redshifts of a few as suggested
by various authors (e.g. Wijers et al. 1997 and ref. therein), 
source evolution may dominate the observed size-frequency distribution, 
and a simple geometrical analysis like this one is not relevant.

\section{Application to GRBs}

In this section we apply our method to GRBs in the Current Catalog 
of BATSE as available on 1997, June 30th (Meegan et al. 1997). 
We have choosen this catalog because it is publicly available,
it contains a large number of bursts, it provides several measures 
of GRB brightness and because BATSE is sensitive enough
to fully sample the homogeneous region.
We compare various measures of the GRB brightness,
like their peak fluxes on different timescales and their fluences
in different energy bands.
We study separately the influence of the time window and of the 
energy range.

\subsection{Short gamma-ray bursts}
Dezalay et al. (1992) and Kouveliotou et al. (1993) have
suggested the existence of two classes of GRBs with different
durations (below and above 2 seconds). 
Although both short and long GRBs are isotropic
on the sky, it is not clear whether they have the same radial distribution 
(the sources of short GRBs could be closer or
farther than the sources of long GRBs).
For this reason we separate the two classes
in our analysis.
The sample of short bursts is unfortunately too small to allow us to draw 
conclusions on their luminosity distribution, and the rest of this paper 
is restricted to the class of long GRBs (1103 bursts).

\subsection{Influence of the time window}
In this section we compare the peak fluxes on 64 and 1024 ms (hereafter 
called P64 and P1), and the fluence in the energy range [50-300 keV]
(hereafter called F23).
These 3 quantities are all defined from 50 to 300~keV but  differ
by the time window used to integrate the burst brightness.

The size-frequency distributions for these 3 parameters are displayed 
in figure~2 (a,b,c), along with the extrapolation of the bright end 
of the distribution.
The deviation from homogeneity occurs at P64~=~6.9~ph~cm$^{-2}$~s$^{-1}$,
P1~=~6.6~ph~cm$^{-2}$~s$^{-1}$ and F23~=~10$^{-5}$~erg~cm$^{-2}$.
This justifies a-posteriori our second assumption, since the catalog
of BATSE is over 99\% complete at and above 
these intensities.\footnote{The trigger efficiency of BATSE is 
directly available in the catalog as a function of P64 and P1. 
For F23 we checked that the burst with the smallest value of P1
in the homogeneous part of the size-frequency curve has P1~=~0.58~ph~cm$^{-2}$~s$^{-1}$, a value above which the Catalog 
is 99\% complete.}
The values of $N_b$ and their 90\% error bars are respectively
129 [78-151], 95 [74-130] and 139 [100-210], for P64, P1 and F23.
These three measures of intensity show little difference
from the point of view of our analysis, suggesting
that the corresponding luminosities 
have comparable intrinsic dispersions.
We conclude that the choice of the time window does not seem to be 
crucial in the search for a standard candle luminosity. 

\subsection{Influence of the energy window}
We now compare the burst fluences in the energy ranges 
[25-100 keV], [50-300 keV] and [100-2000 keV], 
we call these numbers F12, F23 and F34. 
The BATSE catalog does not provide peak fluxes at energies other
than [50-300 keV], so we do not consider peak fluxes in this section.

The size-frequency curves for F12, F23 and F34 are displayed in 
figure~2 (d,e,f), along with the extrapolation of the bright end 
of the distribution.
The departure from homogeneity occurs at F12~=~2.6~10$^{-6}$~erg~cm$^{-2}$,
F23~=~10$^{-5}$~erg~cm$^{-2}$ and F34~=~3.9~10$^{-5}$~erg~cm$^{-2}$.
We again checked that the trigger efficiency of BATSE for GRBs 
in the homogeneous part of the size-frequency is greater than 98\% 
in the 3 cases.
The values of $N_b$ and their 90\% errors are respectively
254 [190-318], 139 [100-210] and 76 [60-112].
These numbers show that the GRB size-frequency distribution contains
about 3 times more GRBs in its homogeneous part 
when F12 instead of F34 is used to measure the burst intensity.
This suggests that the low energy fluence (below
100 keV) is significantly closer to a standard candle 
than the fluence measured at higher energies.
We used the same method to compare the fluences 
in the energy bands F1 [25-50~keV] and F2 [50-100~keV] and found 
very little difference between these two quantities, with $N_b$ being
237 [186-283] and 224 [181-247] for F1 and F2 respectively.
We conclude that F12 provides the closest approximation to a 
standard candle in the Catalog of BATSE and that the energy window 
appears as a crucial parameter in the search for a luminosity 
with a small intrinsic dispersion.

   \begin{figure*}
    \psfig{file=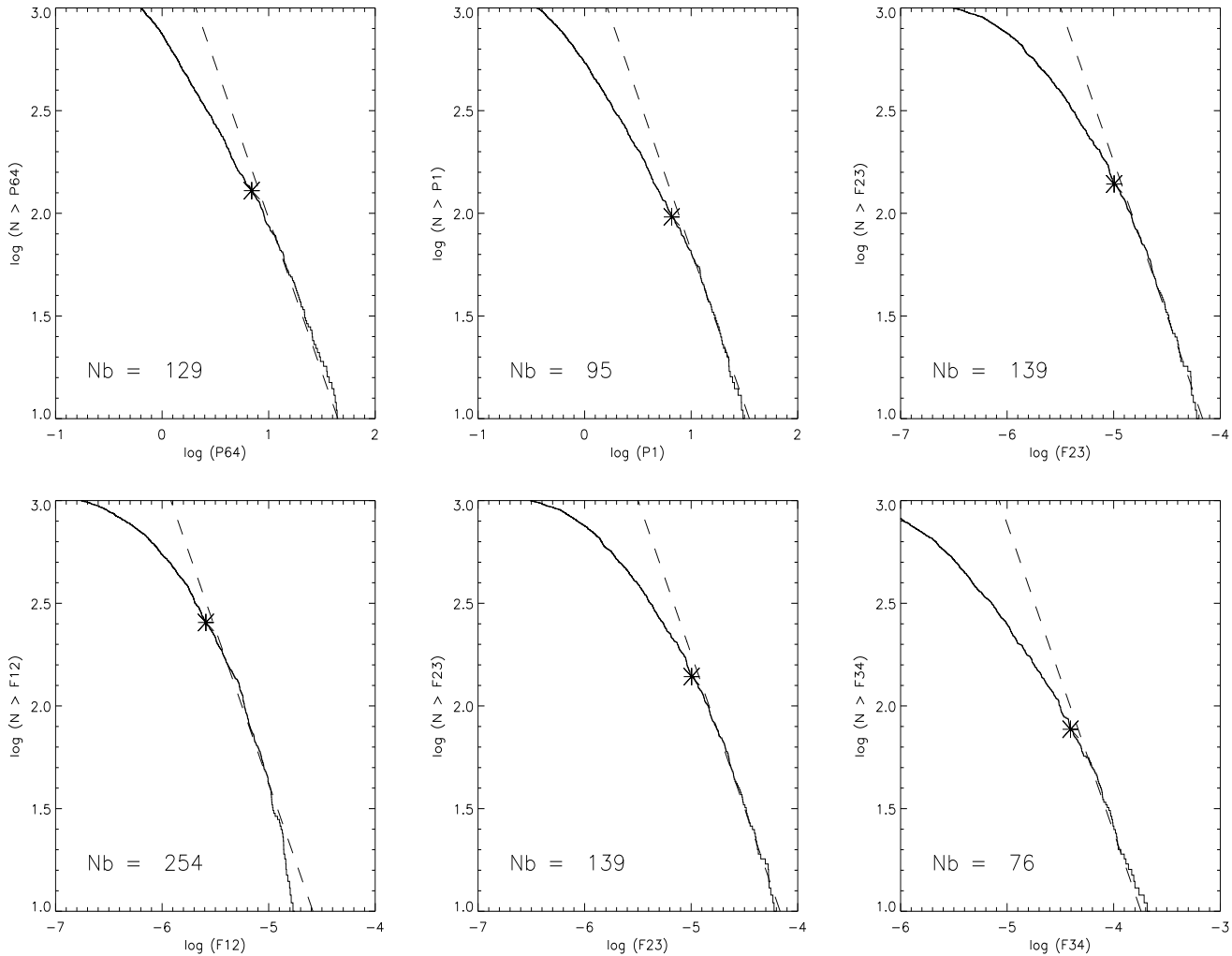,height=14cm,width=18cm}
     \caption{Number of GRBs in the homogeneous part of the 
size-frequency curve for various measures of intensity (see Sect. 3.2).}
         \label{Figsix}
    \end{figure*}

\section{Discussion}
We have proposed a new method to compare the intrinsic dispersion 
of various measures of GRB brightness.
An essential assumption of this method is that the slope -3/2 
at the bright end of the size-frequency curve is due to 
the spatial homogeneity of nearby bursters.
If on the other hand, source evolution dominates the GRB size-frequency
curve, the slope -3/2 might have a completely different interpretation,
invalidating the conclusions of our analysis
(for instance density evolution could cancel the effects of a non-Euclidean
space, simulating a homogeneous distribution of bursters in an 
Euclidean space). 
In the rest of the discussion we assume that source evolution does not 
dominate the observed GRB size-frequency distribution.
 
When applied to GRBs in the Current Catalog of BATSE our method
shows that the measure of luminosity with the smallest intrinsic 
dispersion is the time integrated luminosity below 100~keV.
This result is however not complete since we restricted our analysis
to 5 measures of brightness given in the Catalog. 

This study calls for a careful definition of the GRB brightness when 
this quantity is used as a distance indicator 
(e.g. to compare the properties of nearby and distance bursters); 
measures at low energies are then clearly preferred.

We finally note that the combination of (1) a broad luminosity function 
and (2) a spatial density which varies with the distance 
has interesting consequences for the comparison of faint and bright GRBs.
Intrinsically bright bursts are detected to large distances 
(typically larger than the size of the homogeneous region) where
the burster spatial density decreases rapidly.
Intrinsically faint bursts on the other hand are only visible 
to much smaller distances where the burster density is constant
(if we remain in the homogeneous region) or slowly decreasing.
As a consequence, going to lower intensities increases the number
of bright GRBs much less (in percentage) than the number of 
intrinsically faint bursts. 
In other words, burst classes based on the {\it observed} brightness 
do not contain the same proportion of {\it intrinsically}
bright bursters.
This changing proportion may produce brightness-dependent 
average burst properties (spectral and/or temporal) 
which could strengthen or counteract cosmological effects.
Because our study suggests that the GRB luminosity function is more
extended above 100~keV, we expect GRB properties
to be more brightness-dependent when brightness is measured 
above 100~keV.
For instance the well known hardness-intensity correlation 
(e.g. Dezalay et al. 1997) could be explained in this way 
if it appears that it is stronger when the intensity 
is measured at higher energies.

While we do not address here the reasons which make the luminosity
at low energies a better standard candle,
we note that this behaviour could well be explained by 
an anisotropy of the emission above $\sim$100~keV.
Such an anisotropy would make the brightness at high energies 
dependent on the aspect of the source.
From the point of view of the size-frequency distribution, 
a beaming of the emission is equivalent to the existence
of a luminosity function. 
Hence a beaming factor which changes with the energy 
may just appear as an energy dependent luminosity function, 
which is precisely what we observe.

\begin{acknowledgements}
    The author is grateful to J-P. Dezalay for helpful discussions 
    on the interpretation of size-frequency distributions and to the
    BATSE team for making the BATSE data rapidly available.

\end{acknowledgements}

\end{document}